# JW-ASTClaw: A Generalizable Multi-Agent Framework for Autonomous Solar Telescope and Its Implementation within Chinese Meridian Project

Li-Yue Tong[1,2], Jia-Ben Lin[1,2], Yuan-Yong Deng[1,2,3], Ying-Zi Sun[1,2], Ming-Fu Shao[1,2,3], Hui Wang[1,2,3] and Chen Yang[1,2,3]

[1] National Astronomical Observatories, Chinese Academy of Sciences, Beijing 100101, China; *jiabenlin@bao.ac.cn*

[2] State Key Laboratory of Solar Activity and Space Weather, National Astronomical Observatories, Chinese Academy of Sciences, Beijing, 100101, China

[3] University of Chinese Academy of Sciences, Beijing 100049, China

**Abstract** We present the first deployment of an end-to-end autonomous control system driven by a large language model (LLM) on an operational solar telescope—the Solar Full-disk Multi-layer Magnetograph (SFMM), named JW-ASTClaw. This system employs a multi-agent framework adopting a decoupled three-layer architecture (perception–decision–execution) interconnected through the Model Context Protocol (MCP), which addresses real-time adaptive scheduling under complex environmental conditions while achieving high portability: the perception and decision logic are reused unchanged across instruments, requiring only telescope-specific command interfaces to be adapted. Three perception agents—data-quality-agent, cloud-analyzer-agent, and flare-detector-agent—encode senior observer expertise, including wind jitter detection via limb-ring standard deviation, projected-circle zonal cloud analysis, and multi-band active region identification, as LLM-callable rules, while a central reasoning engine performs multi-source fusion and conflict resolution. The system supports graceful degradation from cloud LLM to local inference and finally to rule-based fallback, designed for remote field stations with unstable connectivity. Cross-season validation on archival data demonstrates 100% cloud detection with zero false positives across 10 distinct observation dates, with active-region counts and positions closely matching the NOAA Solar Region Summary (SRS) reports (102 vs. 100 across 10 separate validation dates). These capabilities significantly improve scientific-intent-driven observation accessibility, enable rapid flare response for space weather monitoring, enhance data usability under adverse conditions, and increase observability during partially cloudy periods. This work represents the first concrete engineering step toward the embodied intelligent solar telescope concept, providing a validated foundation for the transition from automated scheduling to AI-driven autonomous observation.

## 1 INTRODUCTION

Scientific discovery is undergoing a fundamental transformation. Recent breakthroughs in large language models (LLMs) have demonstrated remarkable capabilities in knowledge understanding, multi-step reasoning, and autonomous task planning across diverse domains, from scientific literature synthesis to robotic experiment execution (Brown et al. 2020; Kramer et al. 2023). When integrated with domain-specific sensors and control systems, these models enable a new class of autonomous research instruments capable of real-time adaptive decision-making. In astronomy, Wang et al. (2024) demonstrated LLM-driven automated observation management for a supernova survey, while Lin et al. (2026) articulated the broader vision of embodied intelligent telescopes that understand scientific intent and actively participate in the discovery process.

In solar observation, this transformation is particularly timely: continuous, high-cadence monitoring across multiple wavelengths demands real-time adaptive decisions under rapidly changing environmental conditions—a challenge that has driven decades of automation development yet remains only partially solved. Space missions such as SOHO and SDO have pioneered automated data pipelines with event-triggered observing modes (Domingo et al. 1995; Schou et al. 2012); ground-based networks like GONG implemented standardized quality control for helioseismology (Harvey et al. 1988); SMAT at the Huairou Solar Observing Station achieved high-performance automated observation (Lin et al. 2013); and high-resolution facilities (SST, GST) developed comprehensive data reduction pipelines (de la Cruz Rodr´ıguez et al. 2015; Cao et al. 2010). However, these systems share a fundamental architectural limitation: they operate on deterministic state-machine or priority-queue logic, addressing post-acquisition data processing rather than real-time adaptive scheduling. The challenge of dynamically adjusting observation strategies in response to rapidly changing conditions—patchy cloud cover, variable seeing, or transient solar events—remains largely unsolved for ground-based observatories tasked with both routine synoptic monitoring and real-time space weather alerting.

The Chinese Meridian Project Phase II (CMP-II) establishes a comprehensive ground-based space weather monitoring network across China. The Solar Full-disk Multi-layer Magnetograph (SFMM) at the Ganyu station (118.96°E, 34.995°N) serves as a core solar observation node of the CMP-II, employing four spectral lines—Fe I 5324 Å , H$\beta$ 4861 Å , Ca II 8542 Å , and H$\alpha$ 6563 Å —spanning the photosphere (~ 180–300 km) to the upper chromosphere (~ 1200–1700 km) (Vernazza et al. 1981; Leenaarts et al. 2012). The Fe I 5324 Å channel provides photospheric vector magnetic field measurements through Stokes polarimetry, while H$\beta$ 4861 Å enables Doppler velocity observations. The chromospheric channels Ca II 8542 Å and H$\alpha$ 6563 Å capture plage, filament, and flare ribbon structures at different atmospheric heights. Previous work established the autonomous master control system (MCS) comprising microservices-based subsystems coordinated by a Central Decision System (Tong et al. 2024), demonstrating 2.5-year continuous operation with 15-minute cadence vector magnetic field measurements and 180-second cadence chromospheric observations (Figure 1).

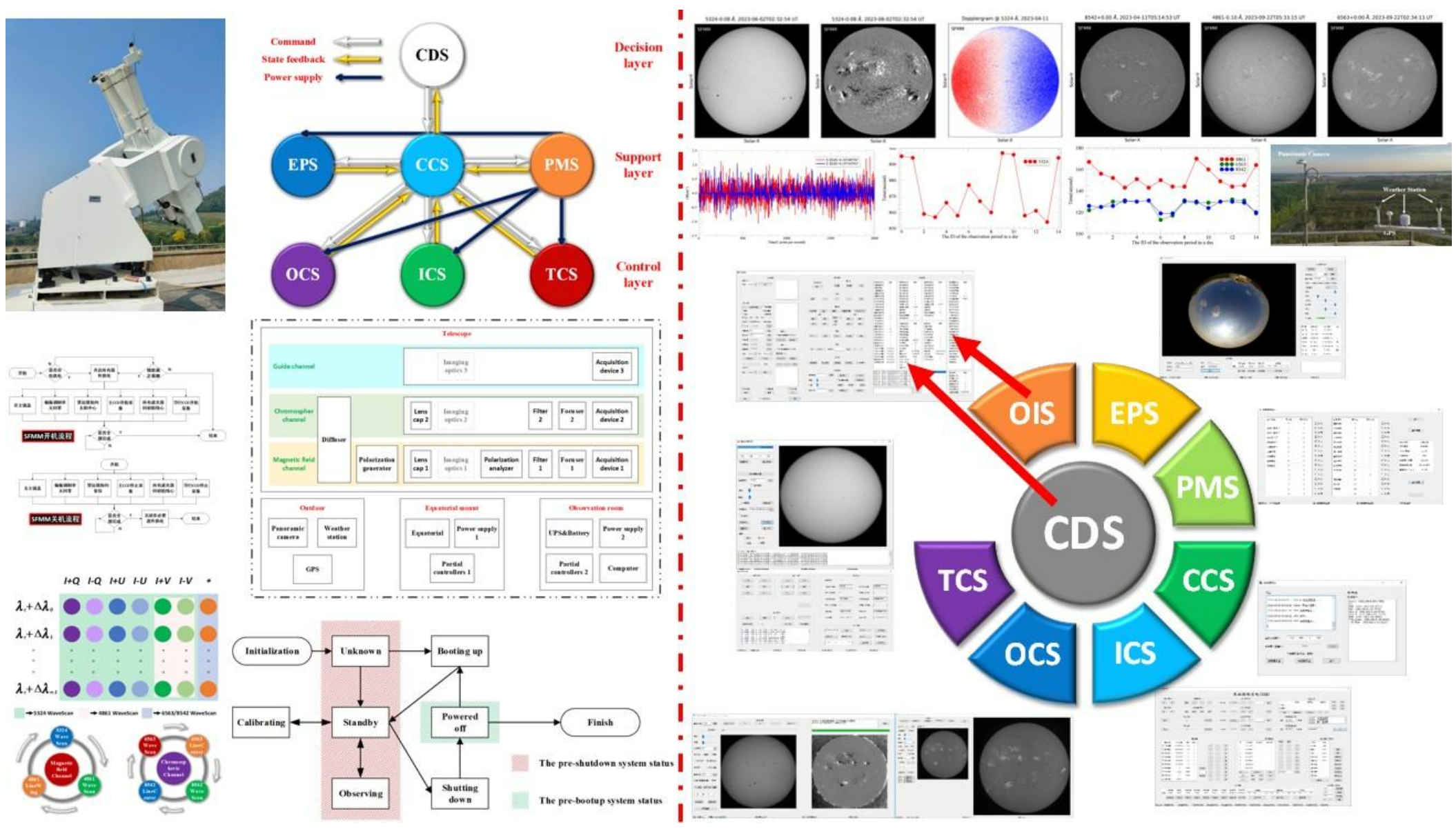

Fig. 1: Overview of the Design and Implementation of SFMM Master Control System.

While this established framework has successfully met routine operational demands, recent breakthroughs in AI-driven reasoning offer a pathway to transition from deterministic scheduling to higher-level cognitive autonomy. To realize this next-generation capability, we identify four functional gaps in the current architecture that constrain its adaptability under complex real-world conditions: (1) *Scientific-intent observation inaccessibility*: custom observation modes require detailed knowledge of the internal instruction set spanning seven subsystems, creating adoption barriers for visiting scientists and limiting the telescope's scientific productivity; (2) *Lack of rapid flare response*: no automated flare detection or observation mode switching exists, despite the space weather monitoring mission requiring immediate response to eruptive events; (3) *Absence of real-time data quality assessment*: no online detection of wind jitter artifacts in magnetograms or cloud occlusion stripe patterns in spectral data, leading to the production of unusable data under adverse conditions; and (4) *Insufficient weather-adaptive scheduling*: the CDS recognizes only binary clear/cloudy states, lacking adaptive scheduling for the partially cloudy conditions frequent at the Ganyu coastal site.

These gaps are particularly acute when multiple adverse conditions coexist. A solar flare occurring during partial cloud cover with moderate wind requires simultaneously balancing weather constraints, scientific priority of the flare event, and data quality implications—a multi-dimensional optimization problem that fixed rule-based systems cannot address, as they can only enumerate predefined scenarios and cannot reason about novel condition combinations. Furthermore, adding new capabilities to rule systems requires modifying existing code, risking disruption of validated functionality.

We address these limitations with JW-ASTClaw, an extensible LLM-driven multi-agent framework built upon a commercial-grade local-first agent runtime and the Model Context Protocol (MCP). To our knowledge, this represents the first end-to-end deployment of an LLM-agent system controlling an operational solar telescope, and the first concrete engineering realization of the embodied intelligent solar telescope concept proposed by Lin et al. (2026). The specific contributions are: (1) *System portability*: a three-layer decoupled architecture (perception–decision–execution) where perception and decision components are reusable across instruments while only telescope-specific execution modules require adaptation, making the "Generalizable" claim in our title concrete; (2) *Algorithmic functionality*: the encoding of senior solar observer expertise into quantitative, LLM-callable rules addressing all four functional gaps—natural language observation control (Gap 1), multi-band active region detection with HDR flare response (Gap 2), wind jitter and cloud stripe quality assessment (Gap 3), and projected-circle zonal cloud analysis with adaptive scheduling (Gap 4)—with cross-seasonally validated thresholds; (3) *System safety*: a six-layer zero-trust safety architecture ensuring that LLM hallucinations or framework failures cannot result in unsafe telescope operations, with full decision traceability; and (4) *Algorithmic robustness*: a dual-LLM graceful degradation design (cloud-hosted reasoning model for normal operation, with architectural provision for local inference fallback) plus four-level failover ensuring the system never operates below the validated rule-based baseline.

## 2 SYSTEM ARCHITECTURE

### 2.1 Three-layer Architecture of JW-ASTClaw

The JW-ASTClaw framework adopts a classical perception-decision-execution architecture (Figure 2), with each layer communicating through the standardized Model Context Protocol (MCP):

**Perception layer** (three intelligent agents): Three domain-specific perception agents provide real-time environmental awareness. The data-quality-agent detects wind jitter, cloud occlusion artifacts, and clarity degradation across all four spectral channels. The cloud-analyzer-agent analyzes panoramic all-sky images for current and approaching cloud conditions. The flare-detector-agent identifies active regions from multi-band observations and monitors them for flare onset. Each agent independently analyzes its data source and generates structured observation advice with confidence scores.

**Decision layer** (JW-ASTClaw reasoning engine): The central orchestrator, developed upon AutoClaw (Zhipu AI)—a commercial local-first agent runtime with built-in guardrails, permission control, and sandboxed execution—integrates advice from all three perception agents and reasons about the optimal observation strategy using the MiniMax M2.7 large language model. It performs multi-source fusion, conflict resolution (e.g., when cloud and flare agents give contradictory recommendations), and priority arbitration based on the current scientific objectives and operational constraints.

**Execution layer** (observation control middleware MCP): The NL-Observer-MCP (Natural Language Observer) translates the decision layer's observation commands into executable telescope instructions, mapping natural language or structured decision outputs to the underlying instrument instruction set spanning seven subsystems.

**Multi-agent collaboration workflow**: The agents collaborate through the following cycle: (1) the three perception agents run in parallel, each analyzing its data source and generating observation advice with confidence scores via MCP; (2) JW-ASTClaw collects all agent reports, performs multi-dimensional reasoning (e.g., cloud-agent reports APPROACHING while flare-agent reports high-probability active region → decision: switch to flare rapid mode with cloud-triggered auto-pause), and issues a unified observation decision; (3) the execution layer translates the decision into telescope commands and confirms execution; (4) the observation outcome is fed back to each perception agent via the `report_decision_feedback` interface, driving memory weight updates and enabling continuous self-calibration. This closed-loop cycle repeats at the cadence of the fastest-changing environmental condition.

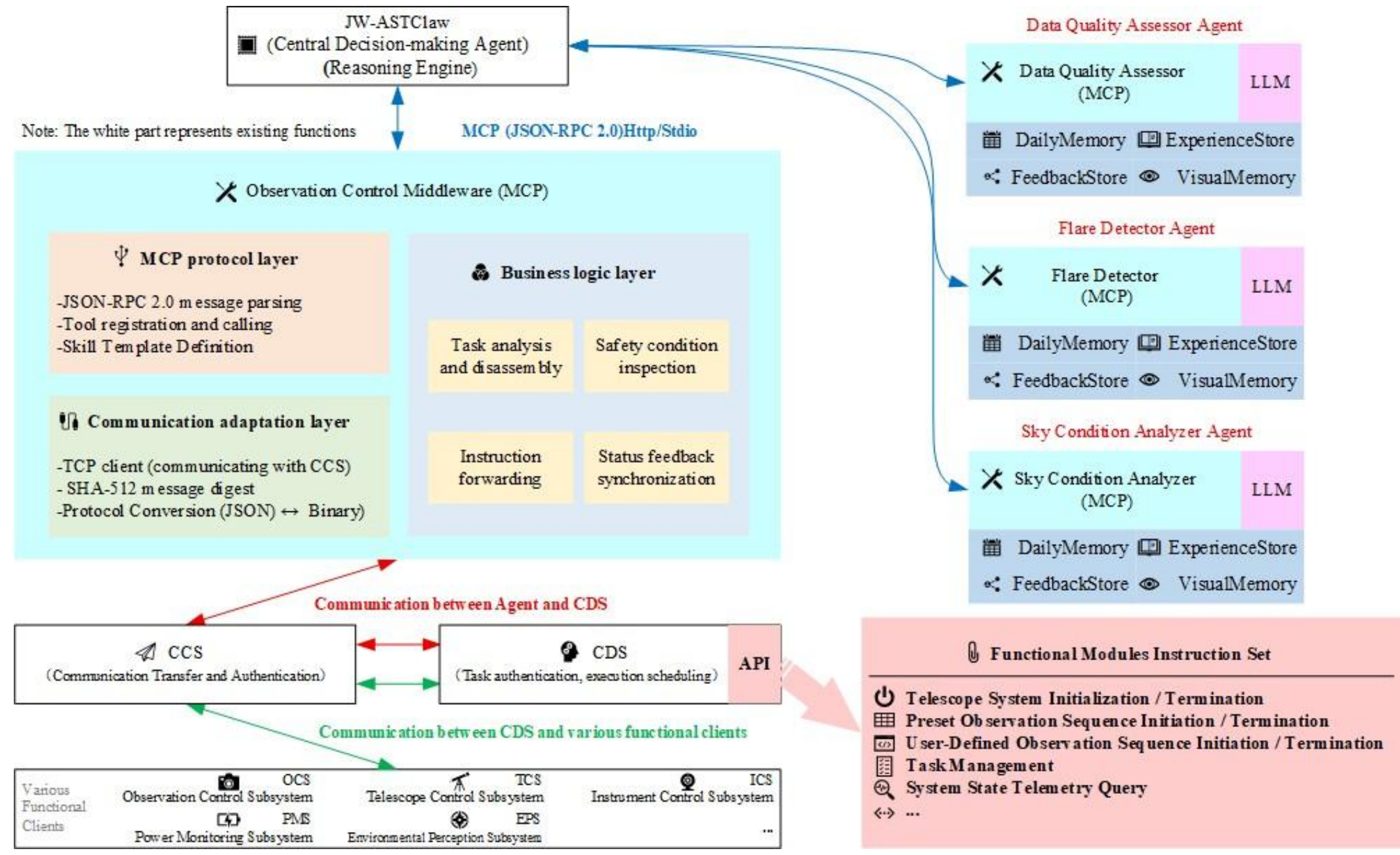

Fig. 2: JW-ASTClaw three-layer architecture. Three perception agents (right) provide environmental awareness; the decision engine (center) performs multi-source reasoning; the execution MCP module (left) translates decisions into telescope commands. All layers communicate through the standardized MCP protocol.

## 2.2 Perception Agent Architecture

Beyond passive tool invocation, each perception module is an intelligent Perception Agent with domain-expert reasoning, structured memory, and experience-driven self-calibration. The agent architecture introduces three capabilities atop the existing analysis algorithms (which remain unmodified):

**LLM-based reasoning**: Each agent possesses a domain-expert role definition (system prompt) and invokes a large language model to synthesize current analysis results with historical experience into actionable observation advice. When the LLM is unavailable, the agent degrades gracefully to a pure data summary response.

**Four-layer structured memory**: (1) *DailyMemory*—an in-memory bounded buffer retaining the current day's analysis events; (2) *ExperienceStore*—a persistent knowledge base with weighted entries and season-priority retrieval; (3) *FeedbackStore*—a record of JW-ASTClaw's adoption/rejection decisions driving weight reinforcement ($w$ += 0.1 × (1 − $w$)) or decay ($w$ ×= 0.85); (4) *VisualMemory*—reference image thumbnails and metadata for visual context during reasoning.

**Nightly consolidation (Dreamer)**: Inspired by sleep-dependent memory consolidation in cognitive science (Park et al. 2023; Sumers et al. 2024), a background scheduler executes three stages each night: (i) daily compression—LLM-summarizes the day's raw events into 3–5 high-quality insights stored in ExperienceStore; (ii) weekly pattern discovery—identifies cross-day trends from recent insights; (iii) conflict resolution—analyzes rejected advice and decays related knowledge weights. This mechanism enables continuous self-calibration without manual intervention.

The feedback loop between JW-ASTClaw and the perception agents creates a closed-loop learning system: agents propose observation advice with confidence scores, JW-ASTClaw makes the final decision and reports the outcome, and agents adjust their knowledge weights accordingly. Over time, each agent accumulates domain-specific experience tuned to the local observing conditions at the Ganyu station.

### 2.3 Execution Layer: Observation Control Middleware

The execution layer bridges the decision engine and the physical telescope through a natural language observation control middleware (NL-Observer-MCP). Its core design comprises three components:

**Instruction set abstraction**: The SFMM instrument control spans seven subsystems (Observation Control Subsystem (OCS), Telescope Control Subsystem (TCS), Instrument Control Subsystem (ICS), Communication Control Subsystem (CCS), Environmental Perception Subsystem (EPS), Power Monitoring Subsystem (PMS), Central Decision-making Subsystem (CDS)) with over 50 low-level commands. The middleware abstracts these into a unified instruction schema defined in YAML, mapping high-level observation intents (e.g., "Stokes scan at 5324 Å ") to sequences of subsystem-specific commands with correct parameter encoding and timing constraints.

**Natural language parsing**: Scientists and the JW-ASTClaw decision engine can issue commands in natural language. The middleware employs LLM-based intent recognition to parse free-form instructions into structured observation parameters (target wavelength, scan range, exposure mode, cadence), then maps them to the instruction schema.

**Safety validation**: Before execution, each command sequence undergoes constraint checking—verifying wavelength ranges, exposure limits, and subsystem state compatibility—to prevent invalid or potentially damaging operations.

### 2.4 Fault Tolerance and Degradation Mechanism

For deployment at remote field stations with variable network connectivity, the system implements a four-level graceful degradation chain: (Level 0) full agent mode with remote LLM reasoning and memory-augmented advice; (Level 1) when network is unavailable, perception agents automatically degrade to passive data-summary mode, producing structured output in $<$ 100 ms without LLM inference, while the decision layer is architecturally designed to fall back to a locally deployed model (Ollama + Qwen3.6-27B), though quantitative benchmarking awaits dedicated GPU hardware deployment at the field station; (Level 2) if the local model also fails, the original rule-based CDS takes over; (Level 3) manual observation as the final fallback.

A HealthMonitor component continuously probes LLM endpoint availability. After three consecutive failures (each with 5-second timeout), the system automatically transitions to Level 1 within one check cycle ($< 60$ s). A single successful LLM call restores Level 0 operation. Mode transitions occur without service restart, and all switching events are logged with timestamps for post-hoc audit.

Importantly, even when the LLM produces plausible but incorrect outputs (e.g., an inappropriate exposure combination), the downstream safety layers (described in Section 2.5) independently validate every command before execution. This design ensures that the system never becomes *less* capable than the pre-existing automation—it can only add intelligence on top of the baseline.

## 2.5 Safety Architecture and Risk Mitigation

### *2.5.1 Platform Security*

JW-ASTClaw is built upon AutoClaw, a commercial-grade local-first agent runtime developed by Zhipu AI. Unlike the original open-source OpenClaw, AutoClaw incorporates enterprise-level security hardening including: built-in execution guardrails that constrain agent actions to declared permission scopes; input/output format validation rejecting malformed or out-of-scope responses; sandboxed tool execution preventing unauthorized system access; and comprehensive audit logging of all agent actions. The reasoning engine uses the MiniMax M2.7 model for multi-source decision-making.

### *2.5.2 Network Security and Deployment*

The SFMM operates on a physically isolated internal network. During the current validation phase, controlled API access through a secured proxy was used solely for rapid capability verification. For production deployment, the system operates in complete physical isolation: a locally deployed LLM (Ollama + Qwen3.6-27B on dedicated GPU hardware) eliminates all external network dependencies. No raw telescope data or observation commands traverse any external network boundary at any stage.

### *2.5.3 Multi-layer Zero-trust Safety Design*

The system employs a six-layer defense-in-depth architecture ensuring that no single point of failure—including LLM hallucinations—can result in unsafe telescope operations:

1. **LLM prompt constraints**: Structured output schemas and domain-restricted system prompts reduce the likelihood of off-domain or malformed responses.
2. **AutoClaw guardrails**: Output format validation rejects responses not conforming to the expected JSON schema.
3. **CDS software safety layer**: The Central Decision Subsystem independently validates every observation command against physical constraints; commands violating safety rules are rejected with error feedback to the reasoning engine.

4. **Subsystem software limits**: Each of the seven control subsystems verifies parameter ranges (wavelength bounds, exposure duration, scan velocity) before execution.
5. **Electronic hardware limits**: Voltage, current, and timing protection circuits prevent out-of-specification electrical operations.
6. **Mechanical limits and thermal fuses**: Physical end-stops, mechanical interlocks, and thermal cutoffs provide ultimate hardware-level protection.

An LLM-generated hallucination command must pass through all six layers to reach the hardware. Any single layer detecting an anomaly terminates execution and logs the event. This zero-trust design ensures safety independent of the AI system's correctness.

#### *2.5.4 Decision Traceability*

Every observation decision is fully auditable. The system logs: (1) all perception agent inputs (analysis data, confidence scores); (2) the complete LLM prompt and raw response; (3) the parsed observation command; (4) CDS validation result (accept/reject with reason); and (5) execution confirmation or error. The FeedbackStore records adoption/rejection outcomes for each piece of advice, enabling post-hoc analysis of decision quality over time.

## 3 CORE MODULE DESIGN & IMPLEMENTATION

The preceding section described the system architecture. This section presents the domain-specific rules and thresholds that constitute the substantive solar-physics contribution of this work—the encoding of senior observer expertise into quantitative, algorithmically executable criteria. For each detection module, we describe: (i) the physical reasoning behind the chosen indicator, (ii) the development dataset used to establish the initial threshold, and (iii) the validated operational threshold after cross-season expansion (Section 4).

### 3.1 Data Quality Real-time Analysis

For magnetogram channels (5324/4861 Å ), the **ring_std** method measures the standard deviation of the normalized magnetic field $10000\times(L-R)/(L+R)$ within a limb annulus ($R$−75 to $R$−20 px). Wind jitter causes position shifts between alternately acquired $P(I+S)$ and $P(I-S)$ polarization frames, producing anomalous fluctuations. During initial development, a threshold of $\sigma_{\rm ring}$ = 12 achieved 94% accuracy on an 18-frame labeled dataset (1 group of 3 Stokes × 6 wavelengths). After cross-season expansion (Section 4), the threshold was refined to $\sigma_{\rm ring}$ = 9.5 for improved sensitivity (Figure 3a, 3b).

For chromospheric integral channels (8542/6563 Å) without polarization data, a **dual-indicator** method compares integral and single-frame ellipse parameters: major axis expansion $d_{\rm maj}$ > 10 px (disk stretched by jitter) or Y/X gradient ratio $g_y/g_x$ < 0.93 (directional motion blur); either indicator exceeding its threshold triggers a jitter detection. Initial validation on a 22-frame development set detected all 7 wind-affected frames with zero false positives. After cross-season expansion (Section 4), thresholds were refined to $d_{\rm maj}$ > 25 px or $g_y/g_x$ < 0.90 (Figure 3c, 3d).

Cloud occlusion in magnetograms produces directional stripe artifacts due to cloud motion between alternately acquired polarization states. These are detected via structure tensor coherence analysis. Initial characterization on a single-day dataset yielded $c \approx 0.116$ (normal) vs. $c \approx 0.063$ (cloud-affected); expanded validation (Section 4) confirmed the threshold at $c = 0.07$ with mean values of 0.075 (normal) and 0.050 (cloud) across 5 groups (Figure 3e, 3f).

### 3.2 Panoramic Cloud Analysis

Traditional all-sky cloud detection methods (HSV, Red-Blue Ratio) proved ineffective at the Ganyu site due to atmospheric haze. We developed a **projected solar disk zonal analysis** method that focuses on the sun's vicinity. The solar disk is expanded 30× (~ 8.1°) and projected onto the all-sky image through a daily auto-calibrated equidistant projection model (adaptive threshold $T = \max(210, T_{\text{Otsu}})$, Nelder-Mead optimization, best-case calibration RMS ~ 7 px; operational acceptance threshold RMS ≤ 25 px). The projected circle is divided into an inner core and 8-sector outer ring aligned with the sun's motion direction. Status: BLOCKED (inner dark > 60%), PARTIAL (30–60%), APPROACHING (ahead > 40%, inner < 20%), CLEAR; the gap [20%, 30%) with low ahead percentage defaults to CLEAR (Figure 4).

### 3.3 Active Region Detection and Flare Monitoring

A note on terminology is warranted. In this work, "active region" (AR) is defined more broadly than the traditional sunspot-group count: our detection pipeline identifies AR candidates based on any combination of (i) photospheric sunspots visible at 5324 Å , (ii) bipolar magnetic field structures in the longitudinal magnetogram, and (iii) chromospheric plage at Ca II 8542 Å . Some detected ARs may lack photospherically visible sunspots—these correspond to NOAA SRS Section II "plage regions without spots." This broader definition is appropriate for the space weather monitoring mission, where magnetically active regions without developed sunspots can still produce flares.

A three-step algorithm exploits SFMM's multi-layer capability. Step 1: photospheric sunspot detection at 5324 Å with limb-darkening-compensated background normalization (ratio < 0.80). Step 2: magnetogram confirmation with bipolar grouping (< 200 px ≈ 145 Mm) and magnetic boundary expansion ($|B| > 80$ G). Step 3: chromospheric plage supplement at Ca II 8542 Å for limb regions ($r/R > 0.85$) where photospheric detection degrades; the chromospheric temperature inversion ensures plage brightening remains visible at the limb. Upon flare detection, the system triggers HDR mode: 6563 Å at 2-second cadence with five-tier bracketed exposures ($n_{\text{exp}}/5$ to $n_{\text{exp}}$, where $n_{\text{exp}}$ is the nominal full exposure) (Figure 5).

### 3.4 Natural Language Observation Control

The natural language interface serves three practical purposes: (1) it enables visiting scientists to configure observations without learning the 50+ low-level commands across seven subsystems; (2) it provides a unified interface for both human operators and the JW-ASTClaw decision engine, eliminating the need to maintain separate command pathways; and (3) it supports rapid on-the-fly adjustments during observation (e.g., "change the scan range from ±0.5 to ±1.0 Å") without interrupting the observation workflow.

The observation control middleware implements a three-stage pipeline for translating high-level observation intents into executable telescope commands. First, the natural language input (from either scientists or the JW-ASTClaw decision engine) is parsed by an LLM into a structured intent representation containing target wavelength, observation mode, scan parameters, and timing constraints. Second, the structured intent is mapped to a sequence of low-level subsystem commands through the YAML-defined instruction schema, which encodes the correct command ordering, parameter encoding formats and inter-command timing requirements for each of the seven subsystems. Third, the generated command sequence undergoes safety validation—checking wavelength range bounds, exposure duration limits, and current subsystem state compatibility—before being transmitted to the telescope via the TCP/IP communication layer. The middleware supports all four observational bands and their associated modes: Stokes polarimetry scans (4861/5324 Å), chromospheric line-center imaging (6563/8542 Å), Doppler velocity field measurement (4861/5324/6563/8542 Å), and the five-tier HDR rapid mode (6563 Å).

When the LLM fails to parse an ambiguous or invalid command, the system returns an explicit error message describing the parsing failure and does not execute any telescope operation. This fail-safe behavior ensures that uncertain inputs never reach the instrument.

## 4 DEPLOYMENT AND VALIDATION

The framework was deployed at the Ganyu station beginning April 2026. Unless otherwise noted, the validation experiments described below were performed in offline replay mode on archived observational data predating the deployment, ensuring reproducibility and enabling cross-season coverage. Table 3 summarizes the validation results across all modules. Table 1 provides a functional capability comparison between the legacy rule-based CDS (operational for 2.5 years) and the JW-ASTClaw framework. Table 2 presents measured end-to-end response latency across all operation modes, addressing the critical requirement for time-sensitive solar event response. The CDS command dispatch itself completes within 50 ms, but the end-to-end latency at Level 2 is bounded by the Modbus protocol communication (~ 500 ms) to downstream subsystem controllers.

Table 1: Functional capability comparison: legacy CDS vs. JW-ASTClaw.

| Capability | Legacy CDS | JW-ASTClaw |
|---|---|---|
| Cloud state assessment | Binary (clear/cloudy) | Four-level real-time status |
| Data quality monitoring | None | Wind jitter + cloud stripe + clarity |
| Active region detection | None | Multi-band three-step algorithm |
| Flare rapid response | None | Auto-detect + HDR trigger |
| Multi-condition reasoning | Not supported | LLM multi-source fusion |
| Natural language interaction | None | Supported |
| Experience accumulation | None | Four-layer structured memory |
| Graceful degradation | None | Four-level fallback |

The data quality module was validated across four seasons (February 2025 to October 2025). For magnetogram wind jitter, the ring_std method was tested on 11 groups (1 original + 10 new covering all seasons; each group nominally comprises 18 magnetograms from 3 Stokes parameters × 6 wavelength points, totaling 176 frames after excluding incomplete scans from groups interrupted by clouds) achieving consistent classification with threshold $\sigma_{\text{ring}} = 9.5$; at this threshold, 29 out of 176 frames were flagged as jitter-affected, matching manual inspection. For chromospheric integral-frame jitter (6563 Å), the dual-indicator method ($d_{\text{maj}}$ > 25 px or $g_y/g_x$ < 0.90) was validated on 10 groups (each containing 11 wavelength-point frames), detecting 39 wind-affected frames consistent with manual assessment. The cloud occlusion stripe detector was validated on 5 groups of magnetograms (each group comprising 18 frames across 3 Stokes × 6 wavelengths, acquired during a dedicated cloudy-day observation on 2026-04-15), with mean coherence of 0.075 for normal and 0.050 for cloud-affected data, correctly separating all cases.

Table 2: Measured end-to-end response latency.

| Scenario | Latency | Description |
|---|---|---|
| NL command → execution (Level 0) | < 5 s | Including LLM inference |
| Perception agent degraded mode (Level 1) | < 100 ms | Agent LLM bypassed; data summary only |
| Rule-based scheduling (Level 2) | < 500 ms | Limited by Modbus protocol |
| Perception agent analysis | < 2 s | Single-frame processing |
| Degradation mode switch | < 60 s | HealthMonitor cycle |

Table 3: Module validation summary.

| Module | Key metric | Result |
|---|---|---|
| Wind jitter (magnetogram) | ring_std accuracy | 29/176 flagged, consistent with manual (11 groups, 4 seasons) |
| Wind jitter (integral) | dual-indicator | 39/110 detected, consistent with manual |
| Cloud (magnetogram) | coherence | 0.075 (clear) / 0.050 (cloud), 5 groups |
| Cloud (panoramic) | detection / FP | 100% / 0% (10 dates, 4 seasons) |
| NL telescope control | parsing accuracy | All commands correct |
| Active region detection | vs NOAA SRS | count ratio 102/100 (10 dates) |

The panoramic cloud analyzer was validated across ten dates spanning June 2025 to April 2026, covering summer (4 dates), autumn (1 date), winter (3 dates), and spring (2 dates) (June 2025 to April 2026): summer (20250612, 20250701, 20250729, 20250829), autumn (20251009), winter (20251210, 20260109, 20260207), and spring (20260320, 20260426). Each date received independent auto-calibration (RMS ≤ 25 px for 9/10 dates). The projected-circle zonal method achieved 100% detection of all cloud occlusion events with zero false positives across all dates. The APPROACHING state correctly identified clouds in outer sectors before they reached the solar position, providing early situational awareness for observation scheduling.

The active region detection algorithm was validated against NOAA Solar Region Summary (SRS) across 10 dates spanning February to December 2025. The three-step method (5324 Å sunspot screening → magnetogram bipolar verification → 8542 Å chromospheric limb supplement) detected a total of 102 active regions compared to NOAA's 100, yielding a count ratio of 102/100. Four out of ten dates achieved exact agreement with NOAA counts, and all dates fell within the ±3 range except for one high-activity date with dense overlapping regions. We note that this count ratio does not constitute a formal precision/recall metric, as one-to-one spatial matching (requiring heliographic coordinate transformation with P-angle correction) was not performed; visual inspection confirmed positional agreement for the matched dates.

To verify spatial agreement quantitatively, we performed P-angle corrected coordinate matching on the four dates with exact count agreement. Spatial matching of 12 AR pairs yielded a mean positional offset of ~ 81″ (~ 8% of the solar disk radius, where 1 pixel ≈ 1″ on the 2048×2048 magnetogram). Decomposition suggests two independent error sources: (i) a baseline ~ $57''$ offset attributable to the difference in center definition between our multi-source fusion centroid and NOAA's sunspot-group centroid (evidenced by the single disk-center match at $57''$), and (ii) an additional ~ $25''$ at the limb from residual P-angle uncertainty in the simplified ephemeris calculation. These offsets are consistent with operational-level agreement without full astrometric calibration. Visual verification of the detection quality is presented in Figure 5.

The HDR rapid observation mode was validated with five-tier bracketed exposures at 6563 Å line center (Figure 6). At the lowest exposure tier ($n/5$), active region fine structure is captured without saturation, while the full exposure ($n$) maintains context of surrounding quiet regions. The five tiers provide continuous dynamic range coverage suitable for capturing both the intense flare ribbon emission and the ambient chromosphere.

The observation control middleware was validated with all test cases covering all seven subsystems and four observational bands, including Stokes scans, chromospheric imaging, Doppler velocity field measurement, and HDR rapid mode. All commands were correctly parsed, mapped, and executed without error.

## 5 DISCUSSION

The JW-ASTClaw framework shifts the paradigm from static rule-based scheduling to memory-augmented agent reasoning. By encapsulating domain-specific perception logic within structured agents, the system resolves the "$M \times N$" integration bottleneck of traditional monolithic controls: new environmental sensors (e.g., seeing monitors or CME alerts) can be deployed as independent agents without modifying the central orchestrator or execution layer. Unlike post-acquisition pipelines (SDO/HMI, GONG) or offline reduction frameworks (SST), our architecture operates in real-time to dynamically balance scientific priority, data quality, and weather constraints—a capability essential for coastal space-weather stations with rapidly varying conditions.

Although the JW-ASTClaw framework marks a paradigm shift in autonomous solar observation, as an emerging architectural paradigm, its practical deployment reveals several constraints that must be addressed. First, the agent experience store requires extended operational cycles before feedback-driven weight adjustments yield measurable improvements in advisory accuracy. Second, the panoramic cloud analyzer currently provides real-time occlusion status but lacks predictive horizon; higher-cadence sampling (10 s) is planned to enable 15–30 minutes advance warning. Third, undetected low-flux active regions currently exhibit negligible flare-eruption potential, though threshold-adaptive detection will be incorporated to expand coverage for marginal zones. Importantly, the four-level graceful degradation chain ensures that agent failures never degrade system capability below the validated rule-based baseline.

Future work will focus on three directions: (1) high-cadence cloud trajectory prediction coupled with agent memory for proactive scheduling; (2) automated algorithm self-evolution, where accumulated reference images and decision outcomes drive periodic model refinement; and (3) cross-wavelength scientific discovery agents that exploit SFMM's simultaneous photospheric–chromospheric observations to identify novel coupling patterns. This work demonstrates a practical pathway from "human-in-the-loop" to "human-on-the-loop" autonomy, applicable to time-domain astronomy and multi-site interferometric networks.

We acknowledge four scope limitations of the current work. (i) *Single-site validation*: the system has only been deployed and validated at SFMM/Ganyu; the "Generalizable" claim in our title is an architectural argument (only the execution layer requires rewriting) that has not yet been demonstrated at a second site. Deployment at Huairou Solar Observing Station is in preparation. (ii) *Threshold calibration sets are relatively small*: the perception agent thresholds (Sections 3.1–3.3) were tuned on the specific datasets stated therein, and overfitting to SFMM-specific local conditions (coastal haze, particular camera characteristics) cannot be entirely ruled out. Cross-site validation will help quantify this risk. (iii) *Local-LLM fallback not yet benchmarked*: while the dual-LLM degradation architecture is designed and implemented, the local model (Qwen3.6-27B) has not been quantitatively compared against the cloud LLM (MiniMax M2.7) due to GPU hardware constraints at the field station; this comparison is planned once dedicated compute hardware is deployed. (iv) *Extreme weather untested*: the perception agents have been validated under normal-to-moderate conditions (haze, partial cloud, moderate wind); heavy snow, dust storms, and strong aerosol events fall outside the current training distribution.

## 6 CONCLUSIONS

This work represents a concrete step in the paradigm shift from automated telescopes to embodied intelligent observatories. By deploying the first LLM-driven multi-agent framework on an operational solar telescope, we demonstrate that the vision of science-intention-driven observation is not merely a conceptual design but an achievable reality. The system demonstrates four key contributions to autonomous solar observation. First, its architecture achieves exceptional portability through a three-layer decoupled design interconnected via the Model Context Protocol (MCP), enabling direct cross-telescope reusability of perception and decision logic without modification. Second, the algorithmic functionality addresses four critical operational gaps through quantitative LLM-callable rules: natural language observation control enhances scientific-intent accessibility for non-expert users; multi-band active region detection with high-dynamic-range flare response enables rapid space weather monitoring; real-time data quality assessment handles wind jitter and cloud stripe artifacts; and projected-circle zonal cloud analysis achieves 100% detection across seasonal conditions, with active-region counts and positions closely matching NOAA SRS reports (102 vs. 100 across 10 validation dates, with a mean positional offset of approximately 81″). Third, system safety is ensured by a six-layer zero-trust architecture that guarantees safe operation independent of AI correctness. Fourth, algorithmic robustness is realized through dual-LLM graceful degradation with four-level failover, ensuring the system never operates below the validated rule-based baseline. The MCP-based modular backbone and experience-driven agent architecture provide a scalable foundation for increasingly autonomous ground-based observatories. By demonstrating that LLM-driven reasoning can reliably operate an operational solar telescope, this work takes a concrete step toward the embodied intelligent observatory paradigm (Lin et al. 2026), where telescopes transition from passive instruments executing preset schedules to active scientific partners capable of understanding intent, reasoning about trade-offs, and learning from experience.

**Acknowledgements** We would like to thank the staff at the National Astronomical Observatories. This work was supported by the National Astronomical Observatories Project of the Chinese Academy of Sciences (No. E4TQ2101), the National Natural Science Foundation of China (NSFC) under Grant No. 11427901, and the Chinese Meridian Project (CMP).

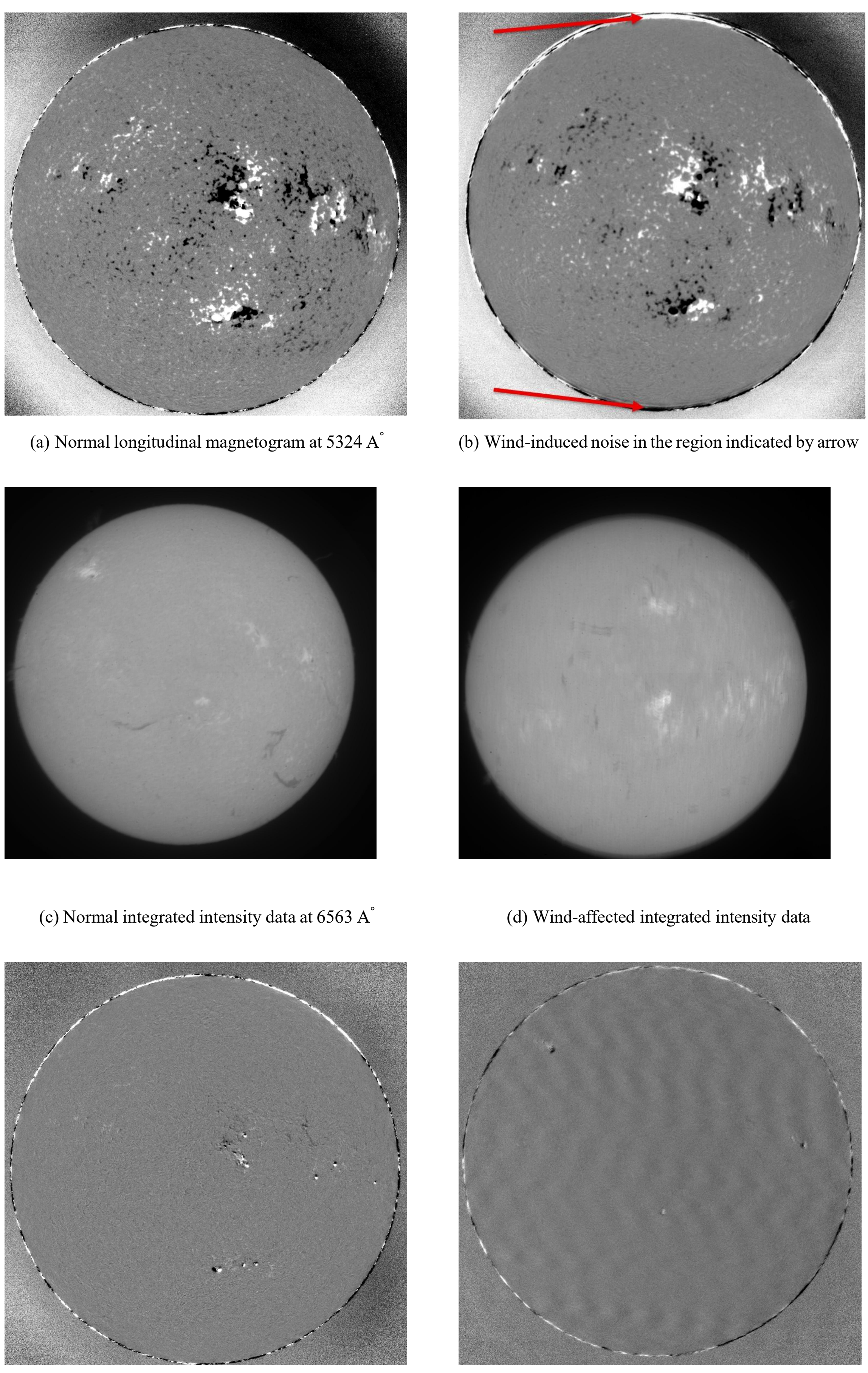

(a) Normal longitudinal magnetogram at 5324 Å

(b) Wind-induced noise in the region indicated by arrow

(c) Normal integrated intensity data at 6563 Å

(d) Wind-affected integrated intensity data

(e) Normal transverse magnetogram at 5324 Å

(f) Magnetogram exhibiting cloud-induced ripples

Fig. 3: Comparison of normal and affected observations.

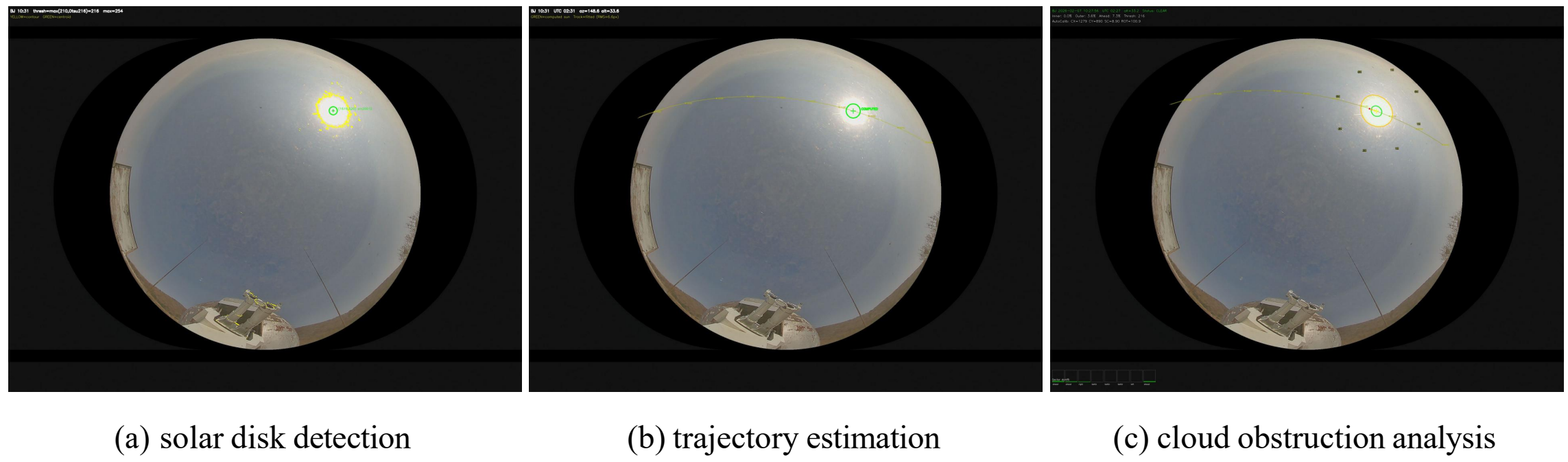

(a) solar disk detection (b) trajectory estimation (c) cloud obstruction analysis

Fig. 4: Three-stage pipeline for solar-region cloud assessment.

(a) Initial sunspot localization at 5324 Å

(b) Magnetic-polarity-constrained region expansion

(c) Augmentation of marginal active regions at 8542 Å

(d) Identification of active regions at 6563 Å

Fig. 5: Multi-wavelength identification pipeline for candidate flare-productive active regions.

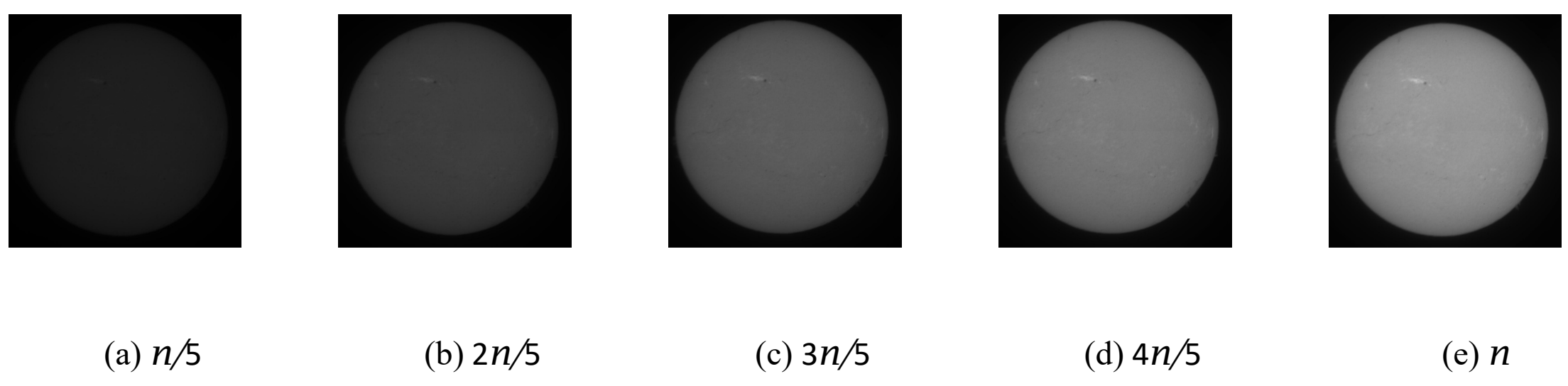

(a) $n/5$ (b) $2n/5$ (c) $3n/5$ (d) $4n/5$ (e) $n$

Fig. 6: Five-tier HDR rapid observation at 6563 Å ($n/5$ to $n$).